\documentclass[aps,reprint,pre,twocolumn,floatfix,amsmath,amssymb]{revtex4}
\usepackage[dvipdfm]{graphicx}
\usepackage{dcolumn}
\usepackage{bm}
\usepackage{color}
\usepackage{amsmath,amssymb}
\usepackage{physics}
\begin{document}
\preprint{}
\title{Heterogeneous Solvent Dissipation Coupled with Particle Rearrangement  in Shear Thinning Non-Brownian Suspensions
}
\author{Tomoharu Terayama} 
\email{terayama@iis.u-tokyo.ac.jp}
\altaffiliation[Present Address:]{ Department of Physics, The University of Tokyo, Hongo 7-3-1, Tokyo 113-0033, Japan.}
\author{Akira Furukawa}
\email{furu@iis.u-tokyo.ac.jp}
\affiliation{Institute of Industrial Science, University of Tokyo, Meguro-ku, Tokyo 153-8505, Japan. }
\date{\today}
\begin{abstract}
Dense non-Brownian suspensions exhibit significant shear thinning, although a comprehensive understanding of the full scope of this phenomenon remains elusive.
This study numerically reveals intimate heterogenous coupled dynamics between many-body particle motions and solvent hydrodynamics in shear-thinning non-Brownian suspensions. We demonstrate the spatially correlated viscous dissipation and particle motions; they share the same characteristic length, which decreases with increasing shear rate. We further show that, at lower shear rates, significant particle density changes are induced against the incompressibility of the solvent, suggesting the cooperative creation and annihilation of gaps and flow channels. We discuss that hydrodynamic interactions may substantially restrict particle rearrangements even in highly dense suspensions, influencing the quantitative aspects of macroscopic rheology. 
\end{abstract}
\maketitle

\section{Introduction}

Understanding and controlling the rheological properties of non-Brownian suspensions, especially the shear-thickening and shear-thinning behavior, are crucial in many industrial applications and are fundamental problems in engineering and basic sciences \cite{MewisB,Stickel05,Denn14,Tanner18,Ness22}. 
The shear-thickening mechanism is currently attracting considerable attention, and recent studies \cite{Seto13,Mari15,Wyart14} proposed a convincing mechanism underlying  the essential role of interparticle contact friction in the steep increase in viscosity. 
In contrast, the shear-thinning mechanism is still under debate \cite{Tanner18,Ness22,Chatte18,Krieger72,Maranzano01,Mari14,Lobry19,Quesada16,Quesada17,Tanner18B,Kroupa17,Quesada18}. 
There are two shear-thinning regimes separated by a shear-thickening regime, and recent rheological experiments combined with mechanical force measurements indicate the following \cite{Chatte18,Comtet17}:  
In the high-shear-rate post-thickening regime, contact friction still dominates the dynamics but is weaker with increasing shear rate, causing shear thinning \cite{Lemaire23}. In contrast, in the low-shear-rate pre-thickening regime, constituent particles are not as strongly compressed against each other such that instead of the frictional contact between particles, the lubrication hydrodynamics should be relevant to particle motion. 
Under such conditions, as the shear rate and the resultant particle pressure increase, the effective particle gaps (volumes) are increased (decreased), reducing the overall flow resistance \cite{Chatte18,Krieger72,Maranzano01,Mari14}.  Although these physical pictures are intuitively appealing, further investigations are required to clarify the physical mechanism of shear thinning at the same level as that of shear thickening. 

When the frictional contact forces are irrelevant to particle motion for lower shear rates, the viscous dissipation due to the solvent dynamics dominates the net dissipation. Furthermore, the incompressibility of the solvent and the resultant hydrodynamic interactions are expected to strongly constrain particle dynamics \cite{Furukawa09,Yamanaka19}. A very recent experimental study certainly indicates that a fluid component (solvent) determines the quantitative aspects of the flow and rheological properties of dense suspensions \cite{NguyenLe23,Lin23}. However, understanding how the solvent hydrodynamics are involved in suspension rheology remains elusive. In this study, using a two-dimensional model non-Brownian suspension, we explore the dynamic coupling between the particle motion and solvent hydrodynamics and its link with the shear-thinning behavior through a direct numerical simulation.

\section{Models and simulation methods}

In this study, our model of a two-dimensional non-Brownian suspension consists of a mixture of large (L) and small (S) disk-shaped particles in a $50:50$ ratio. The size ratio between large and small particles is $1.4$. 
This setup can prevent crystallization, similar to the model system used in previous studies of supercooled liquids \cite{Bernu-Hiwatari-Hansen-Pastore,Yamamoto-Onuki}.   
The $i$-th and $j$-th particles interact via the following very steep soft-core or inverse power law potentials: 
\begin{eqnarray}
U(r_{ij})=\kappa \biggl(\dfrac{s_{ij}}{r_{ij}}\biggr)^{36}\Theta\biggl(\dfrac{r_{ij}}{s_{ij}}-1.4\biggr),  \label{potential}
\end{eqnarray}
where $s_{ij}=(s_{i}+s_{j})/2$, with $s_i$ being the $i$-th particle size, $r_{ij}$ is the distance between the two particles, and $\Theta(r_{ij}/s_{ij}-1.4)$ is the Heaviside step function that provides a cut-off length. 
Here, the soft-core sizes of these particles are denoted as $s_{\rm L}$ and $s_{\rm S}$ for large and small particles, respectively.

Our simulations employ the smoothed-profile method (SPM) \cite{Nakayama05,Molina16,Yamamoto21} to deal with many-body hydrodynamic interactions (HIs) among the constituent particles. 
For this purpose, the $i$-th particle is represented through the following field variables 
$\psi_i({{\mbox{\boldmath$r$}}})$: 
\begin{eqnarray}
\psi_i ({{\mbox{\boldmath$r$}}})=
\begin{cases}
        {1 ~~~~ (|{{\mbox{\boldmath$r$}}}-{{\mbox{\boldmath$R$}}}_{i}|-R < -{\xi}/{2})}\\\\
        {\dfrac{1}{2}\biggl\{1-\sin\biggl[\dfrac{\pi(|{{\mbox{\boldmath$r$}}}-{{\mbox{\boldmath$R$}}}_{i}|-R)}{\xi}\biggr]\biggr\}} \\
        {~~~~~~~~~~
        (-{\xi}/{2} \le |{{\mbox{\boldmath$r$}}}-{{\mbox{\boldmath$R$}}}_{i}|-R \le {\xi}/{2})} \\\\
        {0 ~~~~ (|{{\mbox{\boldmath$r$}}}-{{\mbox{\boldmath$R$}}}_{i}|-R > {\xi}/{2})}
    \end{cases},   
\end{eqnarray} 
where ${{\mbox{\boldmath$R$}}}_{i}$ is the position of the $i$-th particle, $R=s_i/2$,  and $\xi$ is the interface thickness controlling the degree of smoothness. 

The working equations for the velocity field  ${{\mbox{\boldmath$v$}}}({{\mbox{\boldmath$r$}}},t)$ are given as 
\begin{eqnarray}
\rho\biggl(\dfrac{\partial}{\partial t}+ {{\mbox{\boldmath$v$}}}\cdot {{\mbox{\boldmath$\nabla$}}}\biggr){{\mbox{\boldmath$v$}}} &=&  {{\mbox{\boldmath$\nabla$}}}\cdot {\stackrel{\leftrightarrow}{\mbox{\boldmath$\Sigma$}}_{vis}}-{{\mbox{\boldmath$\nabla$}}}p+ {{\mbox{\boldmath$f$}}}_{H}, 
\label{Navier_Stokes} \\ 
{\stackrel{\leftrightarrow}{\mbox{\boldmath$\Sigma$}}_{vis}} &=&  \eta_s \bigl[{{\mbox{\boldmath$\nabla$}}} {{\mbox{\boldmath$v$}}}+ ({{\mbox{\boldmath$\nabla$}}} {{\mbox{\boldmath$v$}}})^T\bigr], \label{stress_tensor} \\
 {{\mbox{\boldmath$\nabla$}}}\cdot {{\mbox{\boldmath$v$}}}&=&0.  \label{incompressibility}
\end{eqnarray}
Equation (\ref{Navier_Stokes}) is the usual Navier-Stokes equation \cite{LandauB}. 
Here, $\rho$ is the mass density and ${\stackrel{\leftrightarrow}{\mbox{\boldmath$\Sigma$}}_{vis}}$ given in Eq. (\ref{stress_tensor}) is the viscous stress tensor, with $\eta_s$ being the solvent viscosity. 
Additionally, the hydrostatic pressure $p$ is determined by the incompressibility condition, Eq.  (\ref{incompressibility}), and ${{\mbox{\boldmath$f$}}}_{H}$ is the body force required to satisfy the rigid body condition \cite{Yamamoto21}.

As described in the main text, periodic boundary conditions are imposed in the $x$-direction with the linear dimension $L$, and planar top and bottom walls are placed at $y=H/2$ and $-H/2$, respectively. Shear flow is imposed by moving the top and bottom walls in the $x$-direction at constant velocities $V/2$ and $-V/2$, respectively, whereby the mean shear rate is given as $\dot\gamma=V/H$. 
We impose no-slip boundary conditions at the top and bottom walls, namely, ${{\mbox{\boldmath$v$}}}(x,y=H/2)=(V/2){\hat {\mbox{\boldmath$x$}}}$ and ${{\mbox{\boldmath$v$}}}(x,y=-H/2)=-(V/2){\hat {\mbox{\boldmath$x$}}}$. 
Furthermore, large and small particles, which are identical to those in the inner region, are embedded in the walls. We let ${{\mbox{\boldmath$R$}}}_{i}^{(w)}=(x_{i}^{(w)},\pm H/2)$ be the position of the $i$-th particle comoving with the walls. 
The wall particles are randomly distributed with the distance of the nearest neighbor pair, $i$ and $j$, being larger than $s_{ij}$ and smaller than $2s_{ij}$. 
This treatment prevents the penetration of the particles through the boundary walls and causes the walls to transmit the forces from the particles.

Regarding the particles that are not on the walls, the equations of motion of the $i$-th particle velocity, ${{\mbox{\boldmath$V$}}}_i$, and the angular velocity, 
${{\mbox{\boldmath$\Omega$}}}_i$, are 
\begin{eqnarray}
M_i \dfrac{d{{\mbox{\boldmath$V$}}}_i}{dt} &=& {{\mbox{\boldmath$F$}}}_{i, {H}} + {{\mbox{\boldmath$F$}}}_{i, {int}}
+ {{\mbox{\boldmath$F$}}}_{i, {ex}},  \label{VG}
\\ 
J_i \dfrac{d{{\mbox{\boldmath$\Omega$}}}_i}{dt} &=&  {{\mbox{\boldmath$N$}}}_{i, {H}} + {{\mbox{\boldmath$N$}}}_{i, {ex}},   \label{OG}
\end{eqnarray}
where $M_{i} = \rho \pi s_{i}^2/4$ and $J_i =M_i s_i^2/8$ are the mass and moment of inertia of the $i$-th particle, respectively. 
In this study, the particle density is assumed to be the same as the solvent density. 
In Eqs. (\ref{VG}) and (\ref{OG}),  ${{\mbox{\boldmath$F$}}}_{i, {int}}$ 
is the force acting on the $i$-th particle due to the particle-particle interactions: 
\begin{eqnarray}
{{\mbox{\boldmath$F$}}}_{i, {int}}= -\sum_{j\ne i} \dfrac{\partial}{\partial {{\mbox{\boldmath$R$}}}_{i}^{}}U(|{{\mbox{\boldmath$R$}}}_{i}^{}-{{\mbox{\boldmath$R$}}}_{j}^{}|),   
\end{eqnarray}
where $U$ is given in Eq. (\ref{potential}).

In Eqs. (\ref{VG}) and (\ref{OG}), 
${{\mbox{\boldmath$F$}}}_{i, {ext}}$ and ${{\mbox{\boldmath$N$}}}_{i, {ext}}$ are the force and torque exerted on the $i$-th particle due to the external field, respectively,  which are absent in the present study. 
Finally, ${{\mbox{\boldmath$F$}}}_{i, {H}}$ and ${{\mbox{\boldmath$N$}}}_{i, {H}}$ are the force and torque exerted on the $i$-th particle due to HIs:
\begin{eqnarray}
{{\mbox{\boldmath$F$}}}_{i, {H}}&=&  -\int {\rm d}{{\mbox{\boldmath$r$}}} {{\mbox{\boldmath$f$}}}_{H},
\label{PFrb}\\
{{\mbox{\boldmath$N$}}}_{i, {H}} &=&  
-\int {\rm d}{{\mbox{\boldmath$r$}}} ({{\mbox{\boldmath$r$}}}-{{\mbox{\boldmath$R$}}}_i) \times {{\mbox{\boldmath$f$}}}_{H}. 
\label{PTrb}
\end{eqnarray}
Equation (\ref{PFrb}) enforces the momentum exchange between the solvent and the particle.  
The body force ${{\mbox{\boldmath$f$}}}_{H}$ can be determined to approximately fulfill the rigid body condition inside the particle region. Its explicit form is given in the discretized equations of motion (for details, please refer to the literature of the SPM \cite{Nakayama05,Molina16,Yamamoto21}). 

In our simulations, we make the equations dimensionless by adopting $\ell$ and $t_0$ as the units for measureing space and time, respectively. Here, $\ell$ represents the discretization mesh size used in solving Eqs. (\ref{Navier_Stokes})-(\ref{incompressibility}), and $t_0=\rho \ell^2/\eta_s$ represents the momentum diffusion time across the unit length. 
As a result, the scaled solvent viscosity is set to $1$, and the units for velocity, stress, force, and energy are defined as $\ell/t_0$, $\rho \ell^2/t_0^2$, $\rho \ell^4/t_0^2$ and $\rho \ell^5/ t_0^2$, respectively. 
In our simulations, we set $\kappa=1$ and $H=L=512$. 
The sizes of the large and small particles are $s_{\rm S}=$8.0 and $s_{\rm L}=$11.2, respectively, and $\xi=1$. 
The area fraction of the particles is defined as 
\begin{eqnarray}
\phi=\dfrac{\pi}{8{\mathcal A}}(2N_{\rm L}s_{\rm L}^2+2N_{\rm S}s_{\rm S}^2+N_{\rm L}^{(w)}s_{\rm L}^2+N_{\rm S}^{(w)}s_{\rm S}^2), 
\end{eqnarray}
where $\mathcal A=H\times L$ represents the area of the system. 
Here, $N_{\rm L}$ and $N_{\rm S}$ represent the number of large and small particles not located on the walls,  respectively, with $N_{\rm L}/N_{\rm S}=1$. 
On the other hand, $N_{\rm L}^{(w)}$ and $N_{\rm S}^{(w)}$ denote the number of large and small particles embedded in the walls, respectively, with $N_{\rm L}^{(w)}/N_{\rm S}^{(w)}=1$. 
Additionally, we ensure that the numbers of particles on the top and bottom walls are identical.

The hydrodynamic interactions (HIs) among disks, satisfying the solvent incompressibility, are incorporated by adopting the smoothed-profile method (SPM) \cite{Nakayama05,Molina16,Yamamoto21}, 
which can accurately reproduce even the near-field HIs \cite{Yamamoto21}:  In the present study, the ratio of the simulation mesh size ($\ell$) to the particle size ($s$) is approximately 0.1, determining the spatial resolution of HIs. With this setting, the SPM can quantitatively replicate the near-field HIs (lubrication interactions) to a satisfactory degree up to a closer distance as $h/s\sim 0.1$ (or even less) \cite{Yamamoto21}, where $h$ represents the gap distance, measured as the separation distance between the interfaces of adjacent particles. 
However, to fully reproduce the singular divergence of lubrication forces, which become crucial at much closer distances of $h/s\lesssim 0.01$, finer resolutions are required. This situation is similar to other hybrid simulation methods \cite{Ladd93,Cates04,Tanaka00,Furukawa18,Espanol95,Jiang11}. 
Nevertheless, in systems of particles with soft or rough surfaces, achieving such conditions is difficult.  
Furthermore, we do not include frictional contact forces in the present model. 
They should be crucial for hard suspended particles at very close contact, while they are expected to be less involved in the actual shear-thinning behavior at lower shear rates \cite{Chatte18,Comtet17}.

\section{Results} 
\subsection{Shear thinning behaviors}

\begin{figure}[htb]
\includegraphics[width=7cm]{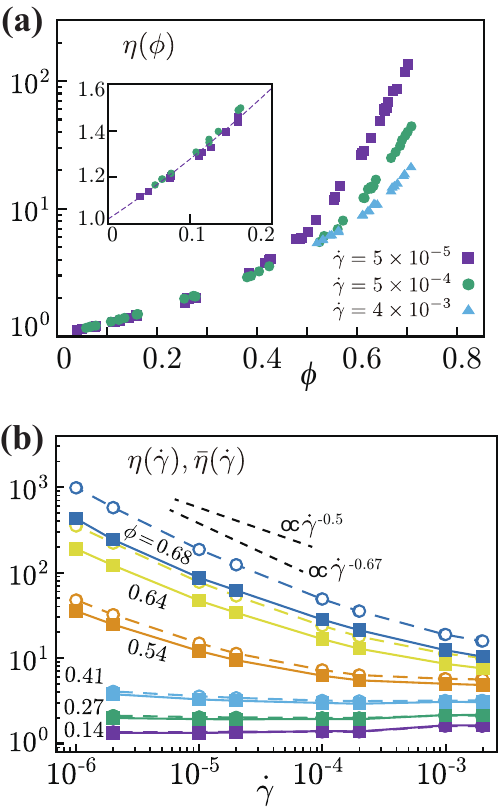}
\caption{(a) $\phi$ dependence of the viscosity $\eta$ for several shear rates $\dot{\gamma}$. The inset shows a closer view for $0 \le \phi \le 0.2$. For $\phi\lesssim 0.4$, $\eta$ almost collapses onto a single curve, while for $\phi\gtrsim 0.4$, $\eta$ increases more steeply for smaller $\dot\gamma$. 
(b) Viscosity {\it vs.} shear rate for various area fractions. The viscosity measured at the walls, $\eta$, is plotted by open circles, while that measured through the dissipation rate, ${\bar \eta}$, is plotted by filled squares. The thinning exponent exhibits a slight dependence on $\phi$ and $\dot\gamma$, and it appears to vary around -0.6 in the thinning regime. 
} 
\label{Fig1}
\end{figure} 

 In this study, the macroscopic viscosity is defined as $\eta=(1/{\dot \gamma}L)\int {\rm d}x \langle \Sigma_{xy}(x,\pm L/2)\rangle$, where $\Sigma_{xy}(x,\pm L/2)$ is the $xy$ component of the stress tensor at the walls and $\langle \cdots \rangle$ hereafter denotes taking the time average in a steady state. 
In Fig. \ref{Fig1}(a), we show the area fraction ($\phi$) dependence of $\eta$ for several shear rates. 
For $\phi\lesssim 0.4$, $\eta$ almost collapses onto a single curve. In contrast, for $\phi\gtrsim 0.4$, $\eta$ increases more steeply for smaller $\dot \gamma$. 
For such higher $\phi$, the force due to the interaction potential predominantly contributes to the net stress, but it increases less linearly with $\dot\gamma$ on average. Therefore, the resultant viscosity is a decreasing function of $\dot\gamma$. 
This situation is more clearly exhibited in Fig. \ref{Fig1}(b), where $\eta$ is plotted against $\dot\gamma$: for the three larger $\phi$ values$(\gtrsim 0.5)$, $\eta$ exhibits marked shear-thinning behavior with a clear plateau at higher shear rates, but we do not find any plateau at lower shear rates in the present ranges of $\dot\gamma$ and $\phi$. 
Note that, in our simulations, the Reynolds number defined as $\rho{\dot \gamma}s_{\rm S}^2/\eta_s$ \cite{Krieger63}, with $\rho$ and $\eta_s$ being the mass density and the solvent viscosity, respectively, spans a range from approximately $10^{-4}$ to $10^{-1}$ for the present range of $\dot\gamma$.

\subsection{Heterogeneous solvent dissipation dynamics}
		
\begin{figure*}[htb]
\centering
\includegraphics[width=13cm]{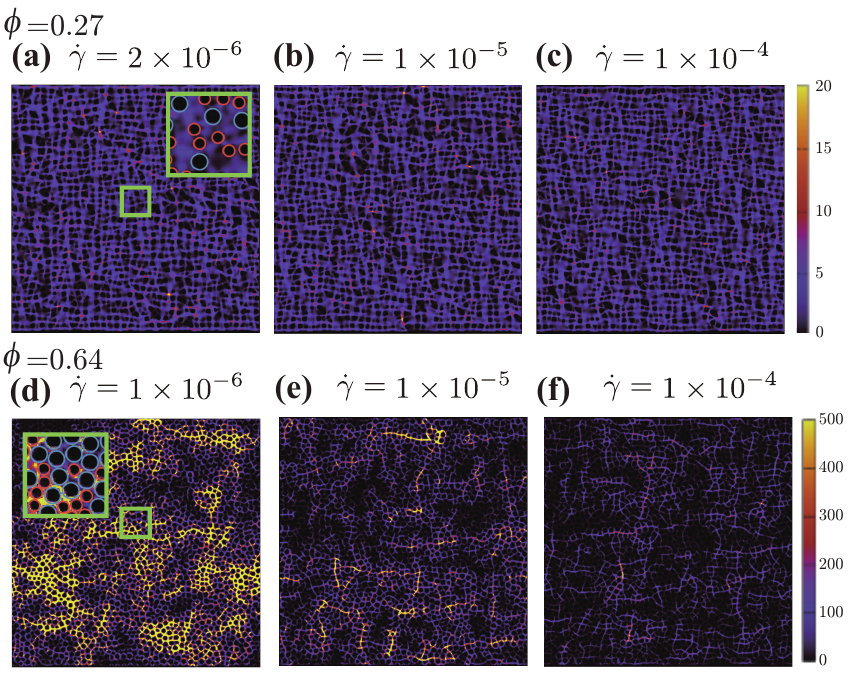}
\caption{(Color online) Typical snapshots of $i_2/\dot\gamma^2$ at $\phi=0.27$ (a)-(c), and those at $\phi=0.64$ (d)-(f) are shown. 
In the insets of (a) and (d), we show the expanded images of small areas. In these insets, although the particles are represented as red and blue circles, we do not distinguish the particle and solvent regions in the main images.  
At $\phi=0.64$, some regions transiently show very steep velocity gradients, whereas those gradients in the other regions are gentle, resulting in the spatially heterogeneous viscous dissipation dynamics. The characteristic size of such heterogeneous dynamics is reduced as $\dot\gamma$ is increased. } 
\label{Fig2}
\end{figure*}

As mentioned above, in the present study, frictional contact forces are not included, and therefore, the dissipation is entirely due to the solvent viscosity.  
The average dissipation rate density, ${\dot \epsilon}_{diss}$, can be written as
\begin{eqnarray}
{\dot \epsilon}_{diss} :=\dfrac{1}{2}{\bar \eta} {\dot\gamma}^2 = \dfrac{1}{\mathcal A}\int_{\mathcal A} \dfrac{1}{2}{\rm d}{ {\mbox{\boldmath$r$}}} \eta_s \langle \bigl(\nabla { {\mbox{\boldmath$v$}}} + \nabla { {\mbox{\boldmath$v$}}}^{\rm T} \bigr)^2 \rangle.   
\end{eqnarray}
Then, the effective viscosity, ${\bar \eta}$, is given by  
\begin{eqnarray}
{\bar \eta}= \dfrac{\eta_s}{{\dot\gamma}^2 {\mathcal A}}\int_{\mathcal A} {\rm d}{ {\mbox{\boldmath$r$}}}  {\langle i_2 \rangle}.   \label{effective_viscosity}
\end{eqnarray}
Here, we introduce the variable $i_2$ as 
\begin{eqnarray}
i_2 &:=& \bigl(\nabla { {\mbox{\boldmath$v$}}} + \nabla { {\mbox{\boldmath$v$}}}^{\rm T} \bigr)^2 \nonumber \\
&=& 2[(\nabla_x v_x -\nabla_y v_y)^2 +(\nabla_x v_y + \nabla_y v_x)^2],   
\end{eqnarray} 
which is positive in the fluid regions and zero in the particle regions. Note that $(i_2/2)^{1/2}$ represents the rotationally invariant shear gradient and $\eta_s i_2/2$ is the local dissipation rate density. In Fig. \ref{Fig1} (b), we plot the effective viscosity $\bar\eta$ defined in Eq. (\ref{effective_viscosity}) in addition to the viscosity measured at the walls, $\eta$. In principle, $\bar\eta$ should coincide with $\eta$, but $\eta$ is significantly larger than $\bar\eta$ at higher $\phi$. Such a deviation between $\eta$ and $\bar\eta$ may be attributed to the intrinsic properties of the SPM: in the SPM framework \cite{Nakayama05,Molina16,Yamamoto21}, the equations of motion are constructed to precisely maintain the momentum exchange between the degrees of freedom of the solvent and particles, but this is not the case for the energy exchange. 
This apparent violation of energy conservation is introduced by the smoothness of the solid-liquid interface of the particles. 
Our preliminary simulations show that when more sharp fluid-particle interfaces are imposed with the particle sizes and total number of particles remaining unchanged, the deviation is significantly reduced (not shown here). Hence, we expect that the {\it quantitative} discrepancy between $\eta$ and $\bar\eta$ does not affect at least the {\it qualitative} aspects of our arguments and results presented below. 

\begin{figure}[htb]
\centering
\includegraphics[width=.7\columnwidth]{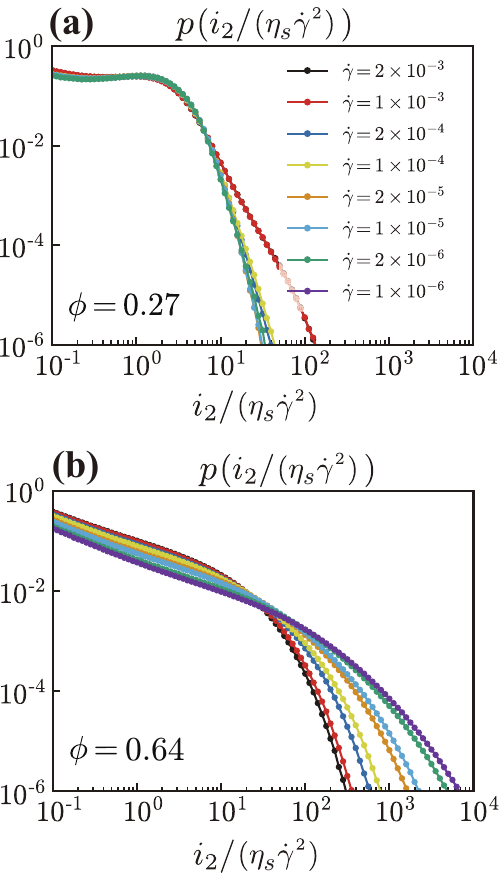}
\caption{(Color online) Probability distribution function of $i_2/(\eta_s \dot\gamma^2)$ at $\phi=0.27$ (a) and $0.64$ (b) for various $\dot\gamma$.  } 
\label{Fig3}
\end{figure}

In Fig. \ref{Fig2}, we plot typical snapshots of the spatial patterns of $i_2/\dot\gamma^2$ for two different $\phi$ with varying $\dot\gamma$. At $\phi=0.27$, where the shear-thinning behavior is hardly observed, the spatial patterns of $i_2/\dot\gamma^2$ remain apparently unchanged. In contrast, at $\phi=0.64$, where marked shear-thinning behavior is observed, we can see a significant $\dot\gamma$ dependence. 
At higher $\dot\gamma(=10^{-4})$, a rather homogeneous spatial pattern of $i_2/\dot\gamma^2$ is observed. However, at lower $\dot\gamma(=10^{-6})$, $i_2/\dot\gamma^2$ is highly heterogeneous, exhibiting a visually clear long-range correlation of much steeper velocity gradients than the average gradient. 
{Although any siginificant anisotropic structures, such as slip or segregation layers and shear bands, are hardly observed near the boundary walls, notable anisotropies in the solvent dynamics are present in the inner region. 
One such anisotropy is observed in $i_2$ around the particles, particularly evident in relatively less dense regimes, as shown in Figs. 2(a)-(c): 
$i_2$ around each particle domain exhibits square-like patterns, which does not indicate structural anisotropy but results from the definition of $i_2$ \cite{comment_anisotropy}. Another observed anisotropy is the string-like patterns in $i_2$, extending horizontally along the $x$-axis and vertically along the $y$-axis, at higher $\dot\gamma$ in dense regimes, as seen in Figs. 2(e) and (f). There are numerous short segments of anisotropic string patterns, characterized by higher values of $i_2$ (in Figs. 2(e) and (f)), which are interpreted as localized manifestations of gliding motions along the shear and flow directions. }

To supplement the observed distinction in the spatial patterns of $i_2$, we calculate the probability distribution of $i_2/(\eta_s\dot\gamma^2)$, $p[i_2/(\eta_s{\dot\gamma^2})]$, as shown in Fig. \ref{Fig3}. 
At $\phi=0.27$, except for at the two higher $\dot\gamma$ values, $p[i_2/(\eta_s{\dot\gamma^2})]$ collapses onto a single curve, which is responsible for the almost flat $\dot\gamma$ dependence of $\eta$.  
In contrast, at $\phi=0.64$, with decreasing $\dot\gamma$, the functional form of $p[i_2/(\eta_s{\dot\gamma^2})]$ becomes broader, with its tail extending to larger values ($i_2/\dot\gamma^2\gg1$). That is, steeper velocity gradients than the applied shear rate $\dot\gamma$ are more enhanced as $\dot\gamma$ is lowered.

\begin{figure}[t]
\centering
\includegraphics[width=.875\columnwidth]{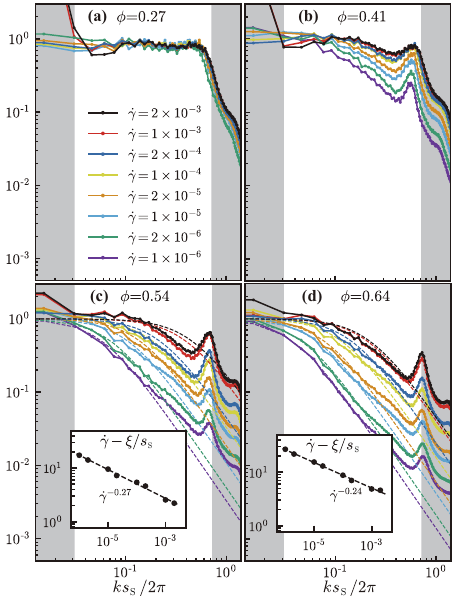}
\caption{Structure factor $S(k)$ of $i_2$ {against $ks_{\rm S}/2\pi$} at $\phi=0.27$ (a), $0.41$ (b), $0.54$ (c), and $0.64$ (d) plotted as solid curves. In our simulations, the available wave numbers are given by $k=2\pi n/L$ with $n$ being an integer.  
For larger $\phi$, (c) and (d), $S(k)$ is fitted to the Ornstein-Zernike form, $S(0)/(1+\xi^2k^2)$ (broken curves), at smaller $k$. 
This fitting determines the correlation length $\xi$, which decreases with increasing  $\dot{\gamma}$. The insets of (c) and (d) show $\xi$ {scaled by the size of the small particle $s_{\rm S}$}. 
 } 
\label{Fig4}
\end{figure}

To further quantify the spatial correlation of the viscous dissipation in the solvent, we define the structure factor of the local velocity gradient as  
\begin{eqnarray}
S( { {\mbox{\boldmath$k$}}} ) =  \dfrac{1}{\mathcal A}\langle |{\hat i}_{2}(\mbox{\boldmath$k$})|^2 \rangle. 
\end{eqnarray}
Here, the Fourier transform of $i_2({\mbox{\boldmath$r$}})$ is defined by 
${\hat i}_{2}(\mbox{\boldmath$k$})=\int d{\mbox{\boldmath$r$}} e^{-i{\mbox{\boldmath$k$}}\cdot{\mbox{\boldmath$r$}}} 
i_2({\mbox{\boldmath$r$}})$. 
Although a window function is not used here, the resultant structure factor for $k>2\times 2\pi/L$ is hardly altered by whether implementing it or not. 
{In the subsequent analysis, our focus is on capturing the trend of decreasing characteristic length scale with an increase in $\dot\gamma$. Despite observing some anisotropies depending on $\phi$ and $\dot\gamma$, replacing $S( { {\mbox{\boldmath$k$}}})$ with its angle average $S(k)$ may not significantly affect the results.} 
Figure \ref{Fig4} displays $S(k)$ for various conditions. 
As expected from the snapshots of $i_2$, $S(k)$ is significantly enhanced at smaller $k$ in the strongly shear-thinning regime, which confirms the spatially correlated viscous dissipation at larger $\phi$ and smaller $\dot\gamma$. Moreover, $S(k)$ at lower $k$ can be fit to the Ornstein-Zernike form 
\begin{eqnarray}
S(k)=\dfrac{S(k\rightarrow 0)}{1+\xi^2 k^2}, 
\end{eqnarray}
from which we can determine the correlation length $\xi$. $\xi$ decreases as $\xi\sim \dot\gamma^{-z}$, with $z\cong 0.25$. 
In recent simulations \cite{Olsson20} for two-dimensional sheared frictionless particles without HIs, a similar exponent was reported for the correlation length extracted from the transverse velocity correlation functions.   

\subsection{Dynamic coupling between particle rearrangement and solvent dissipation}

\begin{figure}[htb]
\centering
\includegraphics[width=0.95\columnwidth]{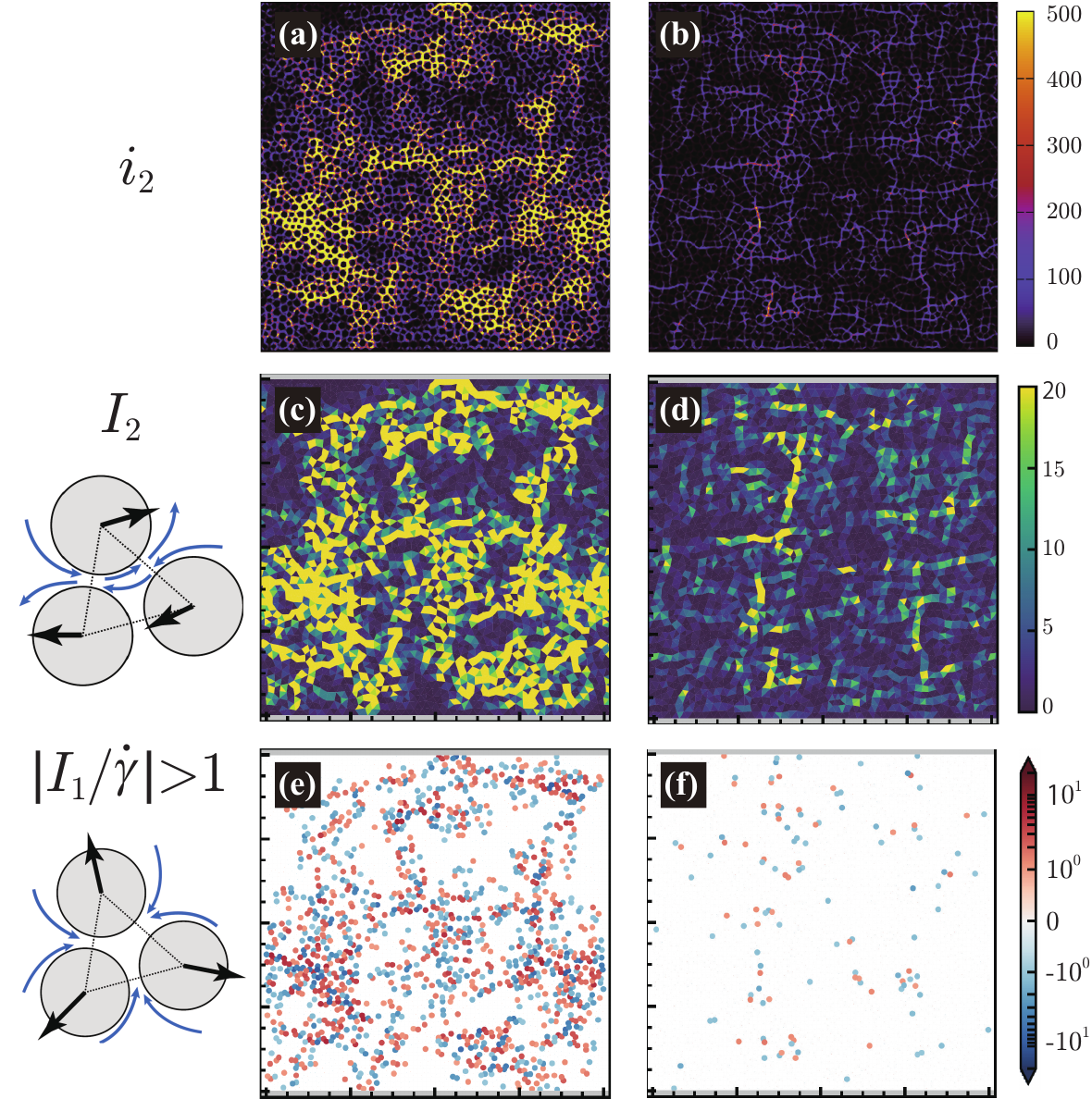}
\caption{Coincidence of the large local dissipation $i_2$ and  relative motion of particles:  
The left and right panels represent typical snapshots of $i_2$ (top), $I_2$ (middle), and $I_1$ (bottom) 
for $\phi=0.64$ at $\dot\gamma=10^{-6}$ and $10^{-4}$, respectively. 
In (e) and (f), points for which $|I_1|/\dot{\gamma}<1$ are omitted.
}
\label{Fig5}
\end{figure}

Let us investigate the link between the observed correlated solvent hydrodynamics and the particle motion. 
To examine this, we introduce the following approach inspired by the finite element method (FEM) \cite{HughesB}.  
Delaunay tessellation (triangulation) is implemented for a sufficiently high-density suspension, with each particle position and velocity being identified with those of a corresponding vertex. 
For the $e$-th triangle, the vertices are labeled as $i=1,2,3$, and the velocity ${\mbox{\boldmath$U$}}_i^{e} = (U_{x,i}^{e}, U_{y,i}^{e})$ is assigned to the $i$-th vertex. 
We quantify the relative particle motion by evaluating the divergence ($I_1^e$) and the rotationally invariant shear gradient ($\sqrt{I_2^e/2}$) of the particle velocities as  
\begin{eqnarray}
I_1^e = {\dfrac{\partial}{\partial x}}\biggl[\sum_{i} N_i^{e}({\mbox{\boldmath$r$}}){U_{x,i}^{e}}\biggr] + {\dfrac{\partial}{\partial y}}\biggl[\sum_{i} N_i^{e}({\mbox{\boldmath$r$}}){U_{y,i}^{e}}\biggr] 
\end{eqnarray}
and
\begin{eqnarray}
I_2^e = ({D_{xx-yy}^e})^2 + ({D_{xy+yx}^e})^2, 
\end{eqnarray}
where 
\begin{eqnarray}
D_{xx-yy}^e &=& {\dfrac{\partial}{\partial x}}\biggl[\sum_{i} N_i^{e}({\mbox{\boldmath$r$}}){U_{x,i}^{e}}\biggr] - {\dfrac{\partial}{\partial y}}\biggl[\sum_{i} N_i^{e}({\mbox{\boldmath$r$}}){U_{y,i}^{e}}\biggr], \nonumber \\
D_{xy+yx}^e &=& {\dfrac{\partial}{\partial x}}\biggl[\sum_{i} N_i^{e}({\mbox{\boldmath$r$}}){U_{y,i}^{e}}\biggr] + {\dfrac{\partial}{\partial y}}\biggl[\sum_{i} N_i^{e}({\mbox{\boldmath$r$}}){U_{x,i}^{e}}\biggr] , \nonumber 
\end{eqnarray}
and $N_i^e({\mbox{\boldmath$r$}})$ is known as the shape function in FEM. Please refer to the SI for details of the expression of $N_i^e({\mbox{\boldmath$r$}})$ and derivations of $I_1^e$ and $I_2^e$.

In Fig. \ref{Fig5}, we plot $I_1$, $I_2$ (based on the particle velocity) and $i_2$ (based on the solvent velocity) at low and high $\dot\gamma$.
The regions with larger values of $I_2$ and $i_2$ are well overlapped, demonstrating that stronger solvent dissipation is associated with larger relative shearing particle motion. 
More interesting is that at such high-shear-gradient locations, significant changes in the local particle density, whose characteristic time scale is faster than that of the applied shear flow ($1/\dot\gamma$), are observed.
At a lower $\dot\gamma$, such local density changes occur cooperatively, and their spatial patterns resemble those of the local shear gradients of the solvent and particle velocities. 
Due to the incompressibility condition of the solvent component, which is completely satisfied in our simulations, the induced solvent flow has a transverse nature (pure shear or rotation). Such a transverse nature of the solvent dynamics should prevent enhancement of the {\it longitudinal} modes of the particle dynamics, whereas we here observe a significant emergence of density changes occurring faster than the time scale of the applied flow ($|I_1|/\dot\gamma \gg1$).  
Note that areas of increases and decreases in particle-density changes are alternatively adjacent to each other with distances of several particle sizes, suppressing the density changes at larger length scales. These behaviors of $I_1$ and $I_2$ are confirmed by evaluating their structure factors shown in the SI. 
Naively, particle rearrangement in suspensions causes simultaneous  flow of the surrounding solvent, and such a forcibly flowing solvent further pushes other particles, in which flow channels of solvents should be formed by creating gaps between particles. For larger $\dot\gamma$, flow channels with sufficiently wide gaps are created. In contrast, for smaller $\dot\gamma$, the shear stress is not sufficiently large, resulting in narrower gaps, by which flow channels or streamlines are extended to almost the same extent. 
In this case, as observed in Fig. 5(e), significant positive and negative particle density changes are alternatively induced in space, indicating that to make flow channels extend over a longer length scale ($\sim\xi$) under narrower average gaps, gap rearrangement against the incompressibility condition is necessary.  

\section{Discussion and Remarks}

We note that shear thinning in non-Brownian suspensions is not solely attributed to HIs; notably, in the Stokes regime of hard-sphere suspensions with a Newtonian solvent, shear thinning does not occur \cite{Krieger63,Lemaitre09,Boyer11}. 
A possible origin of shear thinning could be particle softness or repulsive interactions (effective softness). 
In Ref. \cite{Mari15}, numerical evidence shows that shear thinning becomes more significant with increasing repulsive-interaction range. This finding is consistent with the results of repulsion-range-control experiments \cite{Maranzano01}; especially, when particles can be considered almost hard spheres as the repulsion range is negligibly small, shear thinning was hardly observed \cite{Maranzano01}. 
In stationary states, a larger repulsion range effectively increases particle size, while under shear flow, an increase in the characteristic force magnitude may reduce the effective particle size \cite{Krieger72,Maranzano01,Ness22,Chatte18}. 
This reduction in the particle size can be anisotropic \cite{Furukawa23}, as particle overlaps (gaps) are more enhanced than in stationary states along the compression (extension) axis. 
In our system, HIs with the incompressibility condition should impose strong restrictions on the kinetic paths of particles compared to systems without HIs. 
Such restrictions potentially prevent the particle configuration from undergoing relaxation, subsequently leading to an enhancement of particle overlaps (gaps). In this context, we infer that the involvement of HIs plays a crucial role in determining the quantitative aspects of shear-thinning behaviors. 

The issues mentioned above will be investigated in detail elsewhere by developing comparative simulations with and without HIs. Presently, let us consider phenomenologically how the macroscopic rheology is related to the steady-state properties of solvent flows and the particle configuration as follows. 
The characteristic magnitude of the local shear gradients is set as $\hat{\dot\gamma}$. 
Then, from Eq. (\ref{effective_viscosity}), ${\bar \eta}\dot\gamma^2 \sim \eta_s \hat{\dot\gamma}^2$, and by making use of the relation ${\bar \eta} \sim \dot\gamma^{-\lambda}$, we obtain ${\hat{\dot\gamma}}\sim {\dot\gamma}^{1-\lambda/2}$.  
With $h$ and $s$ being the average ``gap'' and particle sizes, respectively, we may assume that $\hat{\dot\gamma}$ scaled as ${\hat{\dot\gamma}}\sim \dot\gamma s/h$ \cite{Mills09}, where $\dot\gamma s$ is the typical relative velocity between neighboring particles. 
Consequently, we obtain $h\sim \dot\gamma^{\lambda/2}$, indicating that for larger $\dot\gamma$ associated with larger shear stress, the average gap size becomes wider. 
We further assume that the average overlap between particles grows by nearly the same degree as the average gap size and thus that the particle stress almost linearly changes with $h$ \cite{comment_overlap}. 
Therefore, with $G$ being a constant with dimensions of the shear modulus, we may obtain ${\bar \eta} \dot\gamma\sim G h/s$.  Substituting ${\bar \eta}\sim \dot\gamma^{-\lambda}$ and $h\sim\dot\gamma^{\lambda/2}$ into ${\bar \eta} \dot\gamma\sim G h/s$ gives $\lambda\sim 2/3$. This value of $\lambda$ is close to the simulation results for smaller $\dot\gamma$ and larger $\phi$. 

Next, let us set the structural relaxation period as ${\hat \tau}$: 
during ${\hat\tau}$ the average particle configuration undergoes shear deformation with an average shear strain of $\dot\gamma {\hat \tau}$ on the order of $h/s$, and after this period, the configuration may relax. 
Thus, ${\hat \tau}\sim h/\dot\gamma s\sim {\hat{\dot\gamma}^{-1}}$, which immediately means that the time scales of the structural relaxation and interstitial solvent flows are almost identical. 
Moreover, the relation $\bar\eta \dot\gamma\sim G h/s$ can be rewritten as $\bar\eta\sim G h/\dot\gamma s \sim G{\hat{\dot\gamma}^{-1}}\sim G{\hat\tau}$. Namely, the steady-state viscosity is controlled by $h$ or $\hat\tau$, whose values depend on the details of the relaxation mechanism. 
Here, we implicitly assume that the average size of gaps or overlaps, denoted as $h$, is determined through the probability distributions of $i_2$. In the SI, we describe the changes in microstructures with varying $\dot\gamma$ are described for dense case ($\phi=0.64$), but a direct evaluation of $h$ necessitates more careful analysis, including precisely determining the reference when measuring the particle gaps or overlaps.

In this study, we have reported the spatially correlated viscous dissipation in shear-thinning non-Brownian suspensions. Its correlation length $\xi$ decreases with increasing $\dot\gamma$ as $\xi\sim {\dot\gamma}^{-z}$, where $z\cong 0.25$. 
This study has also indicated a close link with the thinning of the viscosity as $\eta\sim \dot\gamma^{-\lambda}\sim \xi^{\lambda/z}$. 
We have further argued that the thinning of $\eta$ is linked to an increase (decrease) in the size of the gaps (overlaps) between particles, {which, however, would not occur in the absence of the particle softness in the first place.} 
At larger $\dot\gamma$, the gap size is greater, by which, to the same extent, the flow channels are shorter and the local velocity gradients are weaker. Such tuning of the local velocity gradients reduces the macroscopic viscosity. 
Then, we have demonstrated that the correlated solvent hydrodynamics are directly coupled with many-body particle motion. 
Intriguingly, significant longitudinal particle motion is induced against the solvent incompressibility over the length scale of $\xi$, which may be associated with cooperative creation of gaps and flow channels. 
Our preliminary simulation results show a significant difference in the observed viscosities with and without HIs in the present ranges of $\dot\gamma$ and $\phi$.  However, this difference seems less pronounced as $\dot\gamma$ decreases and $\phi$ increases. 
More detailed results and discussions of how solvent hydrodynamics influence the macroscopic rheology will be presented elsewhere.

\acknowledgments
We thank Prof. R. Seto for valuable discussions, particularly regarding the significance of the repulsive-interaction range on the qualitative aspects of rheological behaviors. 
This work was supported by KAKENHI (Grants No. 25000002 and 20H05619) and the special fund of Institute of Industrial Science, The University of Tokyo. 

\appendix

\section{Microstructures under shear flow}
{

\begin{figure}[hbt]
\centering
\includegraphics[width=0.9\columnwidth]{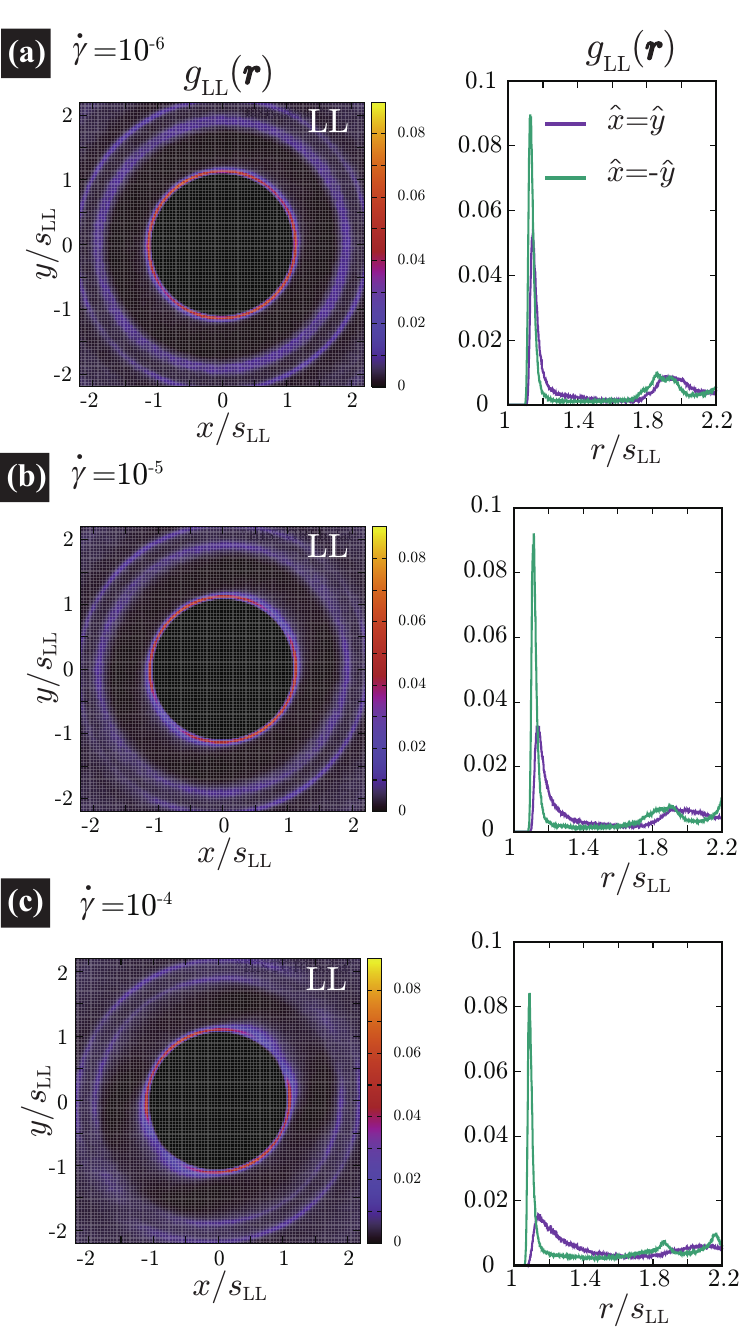}
\caption{$g_{\rm LL}({\mbox{\boldmath$r$}})$ (left) and its cross-sectional  view (right) for $\dot\gamma=10^{-6}$ (a), $10^{-5}$ (b), and $10^{-4}$ (c). 
In the right panels the purple and dark green lines represent $g_{\rm LL}({\mbox{\boldmath$r$}})$ along the extension and compression axes, respectively.}
\label{partial_g}
\end{figure} 

\begin{figure*}[bht]
\centering
\includegraphics[width=1.6\columnwidth]{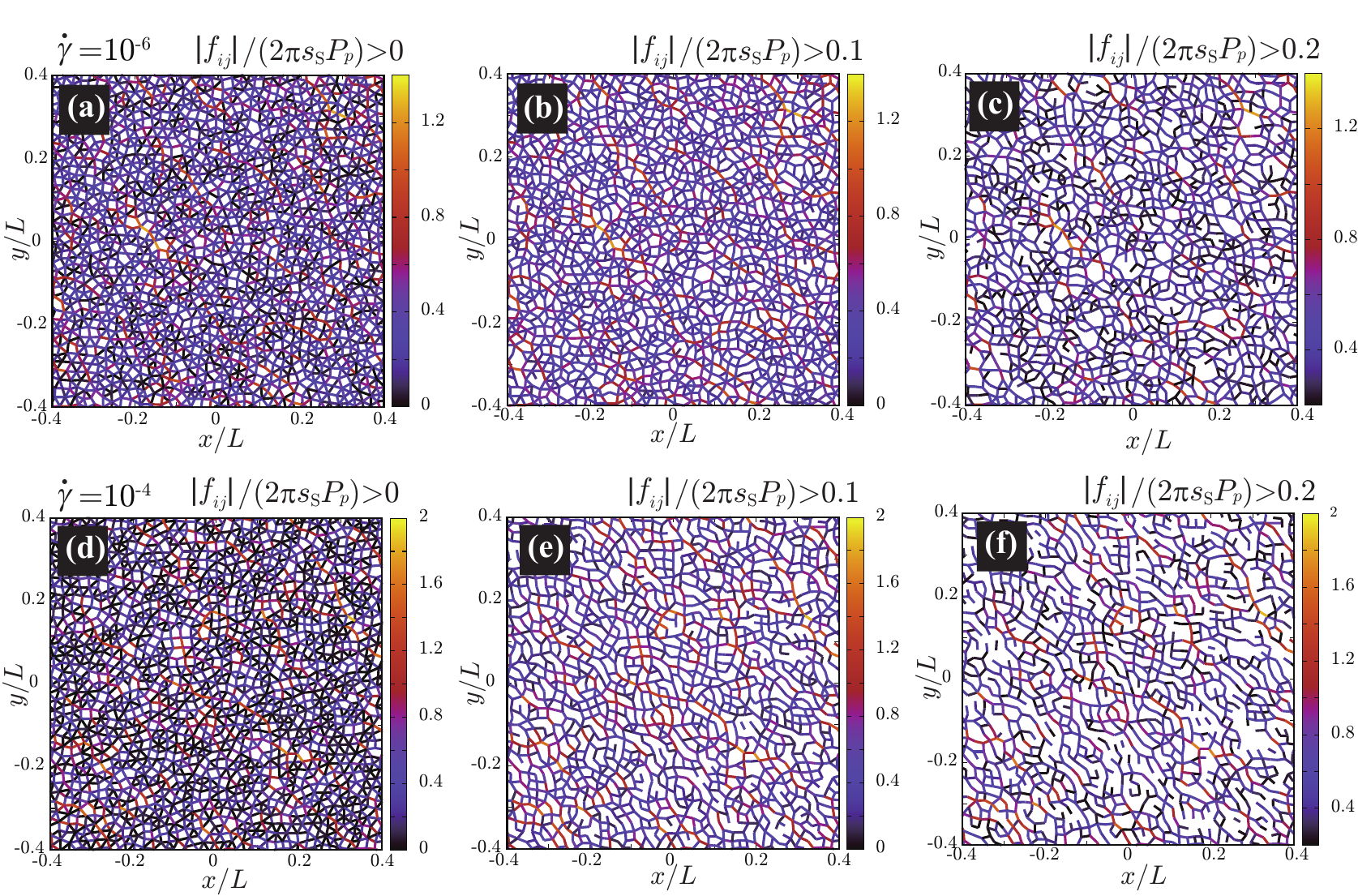}
\caption{Typical snapshots of the force-bonds at $\dot\gamma=10^{-6}$ (a)-(c) and $10^{-4}$ (d)-(f) at $\phi=0.64$ for various threshold values. 
}
\label{force_bond}
\end{figure*} 

In this section,  we examine how the microstructures are modulated by varying the shear rate $\dot\gamma$  
through investigations of the partial pair distribution function (PDF) defined for both the same and different species as   \cite{Hansen_McDonaldB} 
\begin{eqnarray}
g_{\mu\mu}({\mbox{\boldmath$r$}}) = \dfrac{\hat{\mathcal A}}{{\hat N}_\mu ({\hat N}_\mu-1)} {\sum_{i \in \mu}}'{\sum_{j(\ne i)\in \mu}}'\langle \delta({\mbox{\boldmath$r$}}-{\mbox{\boldmath$r$}}_{i}^{(\mu)}+{\mbox{\boldmath$r$}}_{j}^{(\mu)})\rangle,  \nonumber \\ \label{gAA}
\end{eqnarray}
and
\begin{eqnarray}
g_{\rm SL}({\mbox{\boldmath$r$}}) = \langle \dfrac{\hat{\mathcal A}}{{\hat N}_{\rm S} {\hat N}_{\rm L}} {\sum_{i \in \rm S}}'{\sum_{j \in \rm L}}' \delta({\mbox{\boldmath$r$}}-{\mbox{\boldmath$r$}}_{i}^{\rm (S)}+{\mbox{\boldmath$r$}}_{j}^{\rm (L)})\rangle,  \label{gAB}
\end{eqnarray}
respectively.  Here, ${\mbox{\boldmath$r$}}_{i}^{(\mu)}$ represents the position of the $i$-th $\mu$-species particle ($\mu=\rm L,S$).  
In Eqs. (\ref{gAA}) and (\ref{gAB}), the prime denotes that the summantion is taken over the particles for which $-0.3H \le y_{i}^{(\mu)}\le 0.3H$. 
Notice that the walls are located at $y=\pm 0.5H$. 
Additionally, ${\hat N}_{\mu}$ is the average number of $\mu$-species particle in the region of $-0.3L\le y \le 0.3L$ and $\hat{\mathcal A}$ denotes the area of this region given by $\hat{\mathcal A}=0.6 {\mathcal A}$. 

In the left panels of Fig. \ref{partial_g}, we show $g_{\rm LL}({\mbox{\boldmath$r$}})$ at $\phi=0.64$ for various values of $\dot\gamma$ in the shear-thinning regime, demonstrating that a structural anisotropy becomes evident with increasing $\dot\gamma$. 
Along the extension direction (${\hat x}={\hat y}$), the first peak of $g_{\rm LL}({\mbox{\boldmath$r$}})$ becomes smaller and more broad towards the outer region with an increase in $\dot\gamma$, indicating that dilution occurs. 
On the other hand, along the compression axis, almost the opposite occurs. 
Such behaviors are more clearly shown in the right panels of Fig. \ref{partial_g}, showing the cross-sectional views of $g_{\rm LL}({\mbox{\boldmath$r$}})$ along the extension and compression axes. Note that essentially the same behaviors are observed for $g_{\rm SS}({\mbox{\boldmath$r$}})$ and $g_{\rm SL}({\mbox{\boldmath$r$}})$, which are not shown here.

These observations in the partial PDF are supplemented by measurements of the force bond. 
We represent the interaction force magnitude between the $i$- and $j$-th particles as $|f_{ij}|=|{\partial U}/{\partial {{\mbox{\boldmath$r$}}}_{ij}}|$ as the magnitude of the interaction forces between $i$- and $j$-th particles.  
In Figs. \ref{force_bond}(a) and (d), we display typical snapshots of patterns of the force bonds, where the pairs of particles with non-zero interactions ($|f_{ij}>0|$) are described by connecting bonds, for $\dot\gamma=10^{-6}$ and $10^{-4}$, respectively, at $\phi=0.64$. 
Here, the color of each bond indicates the scaled value of $|f_{ij}|$ by $2\pi s_{\rm S} P_p$, with $P_p$ being the particle pressure defined as \cite{Hansen_McDonaldB} 
\begin{eqnarray}
P_p = -\dfrac{1}{2\mathcal A}\sum_i\sum_{j>i} { {\mbox{\boldmath$r$}}}_{ij}\cdot \dfrac{\partial U}{\partial {{\mbox{\boldmath$r$}}}_{ij}}. 
\end{eqnarray} 

For the interaction potential used in this study, as given by Eq. (\ref{potential}), although the potential is steep, the cutoff distance is slightly longer than usual. 
Consequently, when we interpret particle pairs experiencing a non-zero interaction force as being in contact, most particles would be nearly in a perfect contact state for sufficiently large value of $\phi$.  
Nevertheless, this interpretation is oversimplified. 
It is crucial to recognize that the characteristic force magnitude varies with the shear rate. As the shear rate increases, so does the driving external force, leading to a greater characteristic force. Therefore, if the interaction force between particles is significantly smaller than this characteristic force, these pairs can be considered ineffective in contact. 

At this stage, identifying an exact threshold value for judging the interaction force as relevant for contact or not is still difficult.
Instead, by adjusting the threshold value, bonds with force magnitudes below this threshold are not displayed, and subsequently, we assess how their overall patterns change with the shear rate. 
In Figs. \ref{force_bond} (b) and (c), at \(\dot{\gamma} = 10^{-6}\), we display the same snapshots as (a) but with different threshold values.
Furthermore, similar adjustments are made to (d), resulting in (e) and (f).   
A noticeable difference between the cases of $\dot\gamma=10^{-6}$ and $10^{-4}$ is observed even for the threshold values of $|f_{ij}|/(2\pi s_{\rm S} P_p)=0.1$, 
It is observed that more bonds (mainly along the extension axis) disappear at higher shear rates:   
While larger forces are anisotropically exerted along the compression axis, forces along the extension axis are weaker, indicating an effective decrease in contact (or effective opening of gaps) along the elongation axis with an increase in the shear rate. 

Before closing this section, it is noteworthy to mention the following point. 
Although we here scale the interaction forces by $2\pi s_{\rm S} P_p$, the characteristic force magnitude may be alternatively determined by particle shear stress. Notably, a similar trend was found when the force is scaled by the shear stress.

}

\section{Quantifying relative particle motion}

\begin{figure}[h]
\centering
\includegraphics[width=.95\columnwidth]{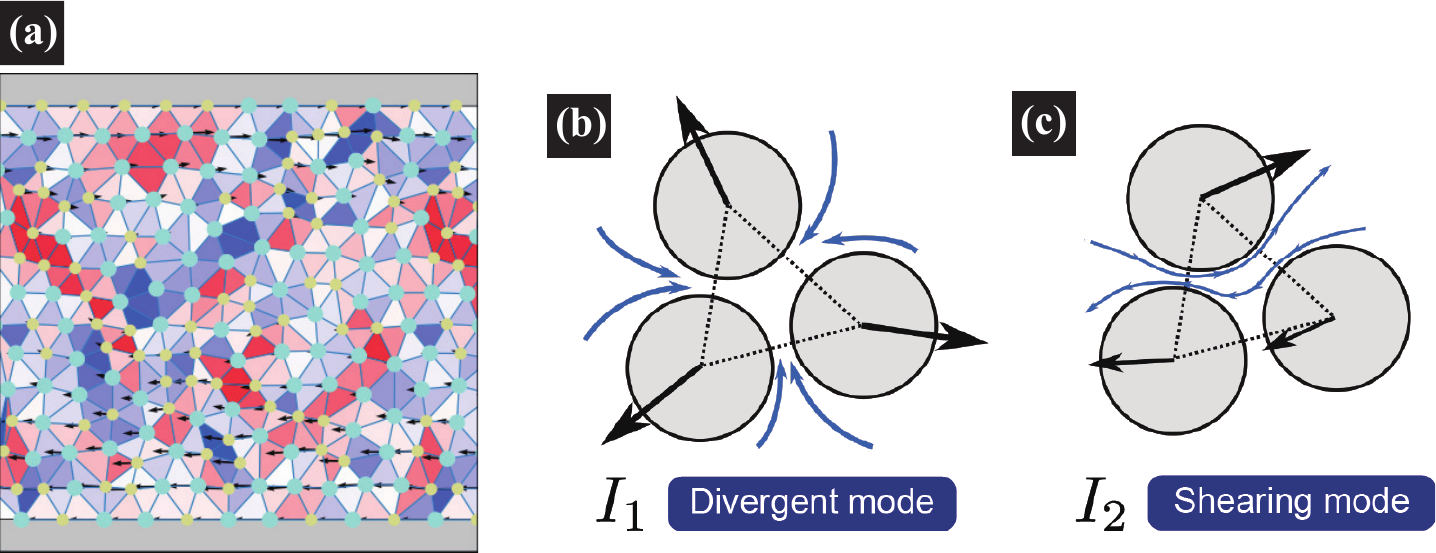}
\caption{(a) Snapshot of $I_1$. Each circle indicates the position of a particle, the color indicates the particle size, and the arrow indicates the velocity. Schematic pictures of $I_1$ (b) and $I_2$ (c). $I_1$ and $I_2$ measure the degree of divergent and shearing motions of the triangular element, respectively. The blue arrows schematically illustrate typical accompanying flows of the surrounding solvent.}\label{Fig::example}
\end{figure} 

We explain the details of our analysis of relative particle motion. To quantify the relative motion, we define two quantities, $I_1$ and $I_2$, whose physical meanings are schematically shown in Fig. \ref{Fig::example}: $I_1$ directly measures the rate of change of the triangular area, while $I_2$ measures the degree of shear deformation rate of the triangle. These quantities are calculated for each element. 
We perform Delaunay triangulation, in which the vertices are the positions of the particles ${\mbox{\boldmath$R$}}_i$. 
The solid lines in the left panel of Fig. \ref{Fig::example} show the partitioning. 
The quantities $I_1$ and $I_2$ are tied to each triangular element. In this way, we can regard $I_1$ and $I_2$ as spatial variables. 

\begin{figure}[htb]
\centering
\includegraphics[width=0.925\columnwidth]{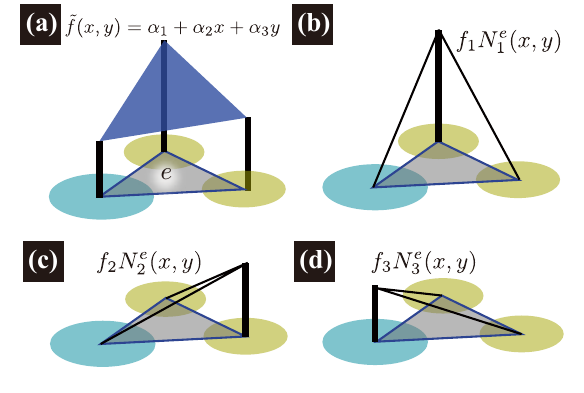}
\caption{(Left) A first-order triangular element.  On each element, the approximate function ${\tilde f}(x,y)$ is described as a plane (a). The plane function ${\tilde f}(x,y)$ can be expressed as a product of shape functions $N_i^e(x,y)$ within the element and the values at the vertices $f_i^e$ (b-d). 
}
\label{Fig::elements}
\end{figure} 

Here, we define the two quantities $I_1$ and $I_2$ as invariants of a tensor. 
The derivation follows the simple procedure of the conventional finite element method (FEM) \cite{HughesB}. We pick up a triangular element $e$, whose vertices are numbered as $1,2$, and $3$ (see Fig. \ref{Fig::elements}).  
We let $f^e_i$ denote the value of an arbitrary variable ${\tilde f}$ at the vertex $i$ ($i=1,2,3$) on the element $e$. 
The vertex positions are denoted as $(x^e_1, y^e_1)$, $(x^e_2,y^e_2)$, and $(x^e_3,y^e_3)$. 
Then, we formally express $f_i^e$ as 
\begin{align}
\mqty(f^e_1\\f^e_2\\f^e_3) &= \mqty(1 & x^e_1 & y^e_1 \\ 1&x^e_2&y^e_2 \\ 1&x^e_3&y^e_3)\mqty(\alpha^e_1\\\alpha^e_2\\\alpha^e_3) =: S\mqty(\alpha^e_1\\\alpha^e_2\\\alpha^e_3), 
\end{align}
where the parameters $\mqty(\alpha^e_1&\alpha^e_2&\alpha^e_3)$ are given by  
\begin{align}
\mqty(\alpha^e_1\\\alpha^e_2\\\alpha^e_3)
= S^{-1}\mqty(f^e_1\\f^e_2\\f^e_3) 
\end{align}
with
\begin{eqnarray}
S^{-1} &=&  \mqty( a^e_1 & a^e_2 & a^e_3 \\ b^e_1 & b^e_2 & b^e_3 \\ c^e_1 & c^e_2 & c^e_3 )  \nonumber \\ 
&:=&\frac{1}{2A_e} \mqty( x^e_2y^e_3 - x^e_3y^e_2& x^e_3y^e_1 - x^e_1y^e_3&x^e_1y^e_2 - x^e_2y^e_1\\
y^e_2 - y^e_3&y^e_3 - y^e_1&y^e_1 - y^e_2\\
x^e_3 - x^e_2&x^e_1 - x^e_3&x^e_3 - x^e_2), \nonumber\\
\end{eqnarray}
where $A_e=\det(S)/2$. 
In the linear FEM, by defining the shape function as $N^e_i(x,y) = a^e_i + b^e_ix + c^e_iy\quad(i=1,2,3)$, we express the field variable ${\tilde f}$ on the element $e$ by interpolating the values assigned to the vertices as   
\begin{align}
\tilde{f}(x,y) &= \sum_{i=1,2,3} N_i^e({{\mbox{\boldmath$x$}}})f_i^e \nonumber \\
&=\mqty(N^e_1(x,y) & N^e_2(x,y) & N^e_3(x,y))\mqty(f^e_1\\f^e_2\\f^e_3),   \label{f_tilde}
\end{align}
where ${{\mbox{\boldmath$x$}}}=(x,y)\in e$.

Let $(V_{xi}^e, V_{yi}^e)$ be the velocity of the $i$-th particle on the element $e$.  
According to the above explained procedure, we can approximate the particle velocity `'field'' at an arbitrary position ${{\mbox{\boldmath$x$}}}$ on the element $e$ as the linear interpolation of $(V_{xi}^e, V_{yi}^e)$: 
\begin{eqnarray}
V_x^e({{\mbox{\boldmath$x$}}})&=& \sum_{i=1,2,3} N_i^e({{\mbox{\boldmath$x$}}})V_{xi}^e,  \\
V_y^e({{\mbox{\boldmath$x$}}})&=& \sum_{i=1,2,3} N_i^e({{\mbox{\boldmath$x$}}})V_{yi}^e. 
\end{eqnarray}
Then, we can compute the deformation rate as 
\begin{eqnarray} \mqty(D_{xx}^e & D_{xy}^e \\ D_{yx}^e & D_{yy}^e ) := \mqty(\frac{\partial}{\partial x}{{V}_x^e({{\mbox{\boldmath$x$}}})} &\frac{\partial}{\partial x}{{V}_y^e({{\mbox{\boldmath$x$}}})}\\ \frac{\partial}{\partial y}{{V}_x^e({{\mbox{\boldmath$x$}}})} &\frac{\partial}{\partial y}{{V}_y^e({{\mbox{\boldmath$x$}}})}), 
\end{eqnarray}
which further defines 
\begin{eqnarray}
I_1^e &:=& D_{xx}^e + D_{yy}^e,  \\
I_2^e &:=& (D_{xx}^e - D_{yy}^e)^2 + (D_{xy}^e + D_{yx}^e)^2. 
\end{eqnarray}

\begin{figure}[htb]
\centering
\includegraphics[width=0.985\columnwidth]{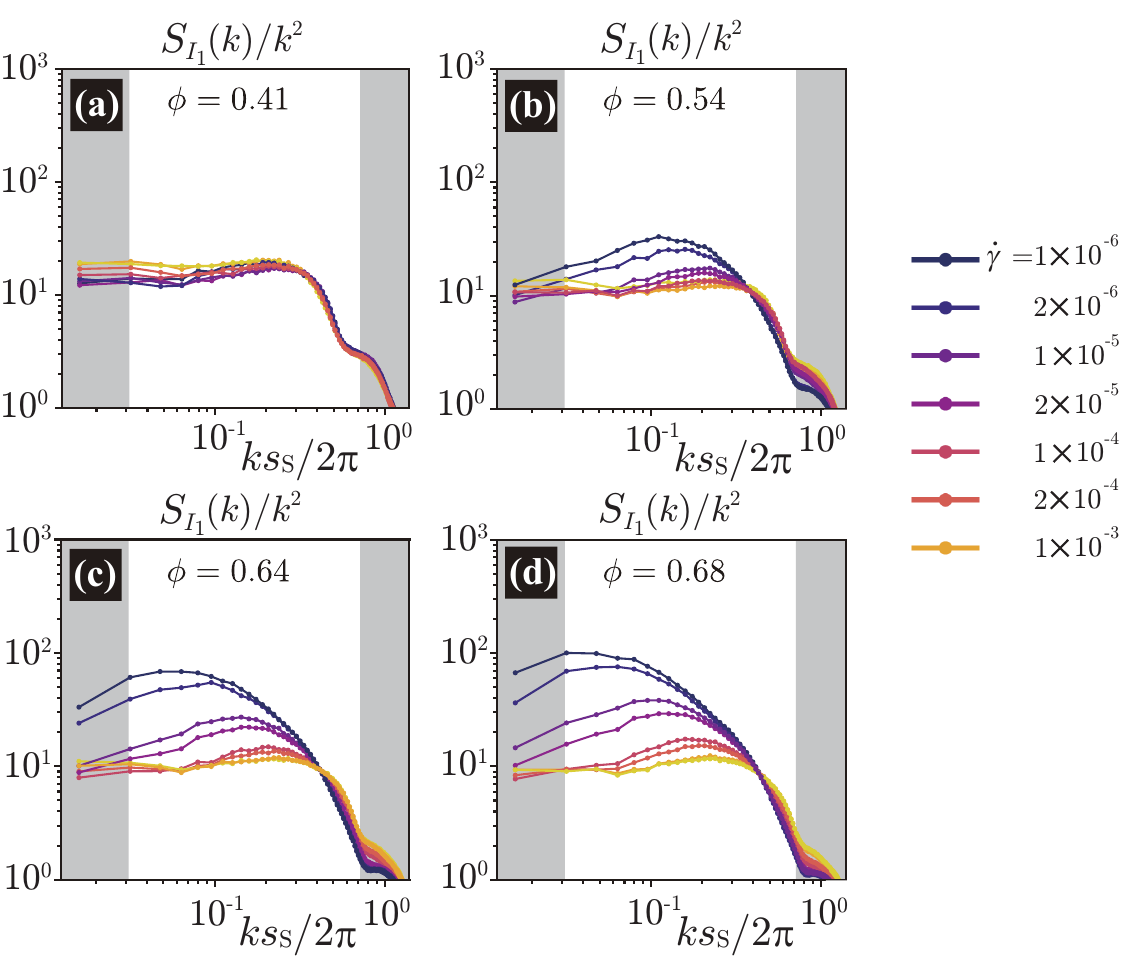}
\caption{ $S_{I_1}(k)/k^2$ {against $ks_{\rm S}/2\pi$} at $\phi=0.41$ (a), $0.54$ (b), $0.64$ (c), and $0.68$ (d).  
At a lower $k$, enhancement of $S_{I_1}(k)$ is suppressed and behaves as $k^a$ ($a\sim 2$). 
}
\label{FigI1}
\end{figure} 

\begin{figure}[htb]
\centering
\includegraphics[width=0.985\columnwidth]{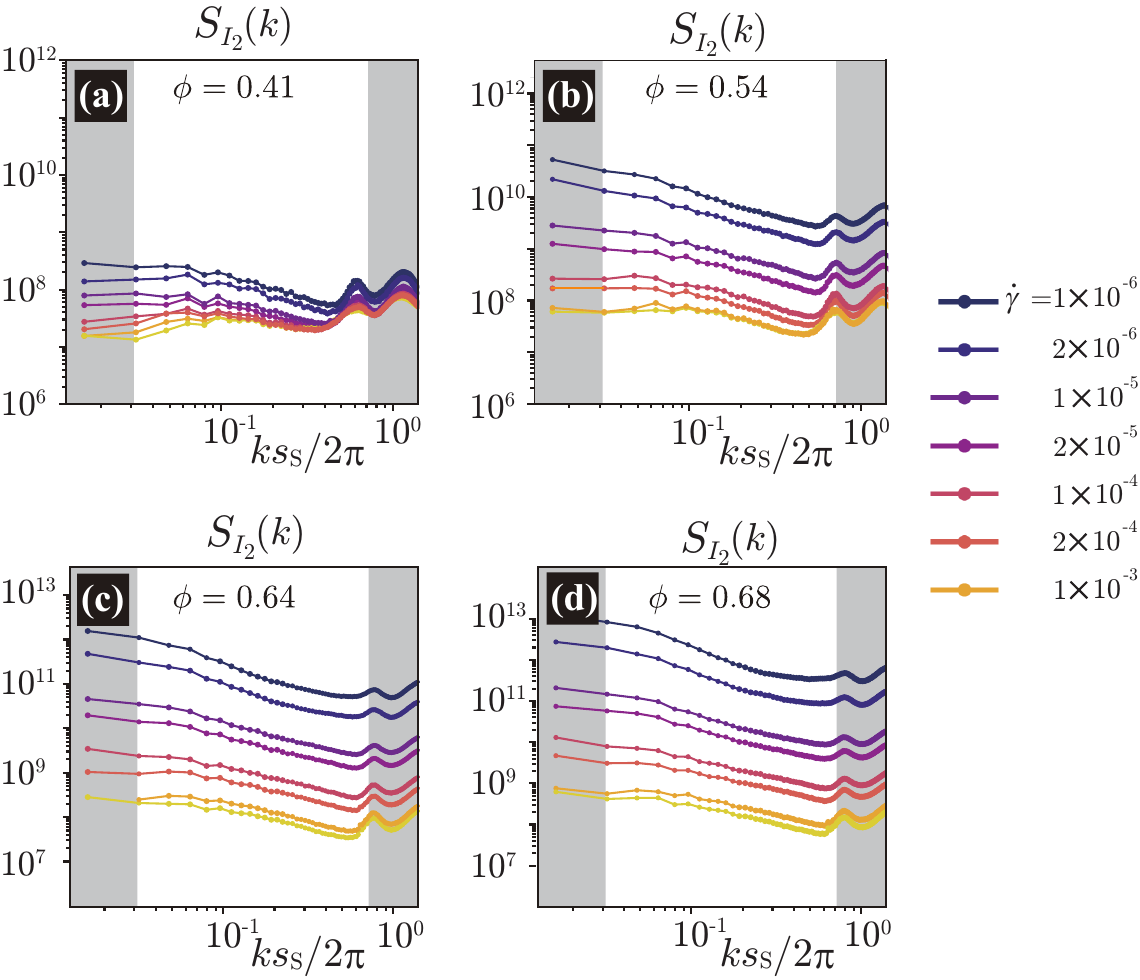}
\caption{
Structure factor $S_{I_2}(k)$ of $I_2$ {against $ks_{\rm S}/2\pi$} at $\phi=0.41$ (a), $0.54$ (b), $0.64$ (c), and $0.68$ (d). 
For larger $\phi$,  $S_{I_2}(k)$ at lower $k$ grows with decreasing $\dot\gamma$. 
Such behaviors are qualitatively similar to those of the structure factor of $i_2$ shown in Fig. 4 in the main text.  
}\label{FigI2}
\end{figure} 

To quantify the spatial correlation of the transverse and longitudinal particle motions,  we define the following structure factors:   
\begin{eqnarray}
S_{I_1}( { {\mbox{\boldmath$k$}}} ) &=&  \dfrac{1}{\mathcal A}\langle |{\hat I}_{1}(\mbox{\boldmath$k$})|^2 \rangle,\\  
S_{I_2}( { {\mbox{\boldmath$k$}}} ) &=&  \dfrac{1}{\mathcal A}\langle |{\hat I}_{2}(\mbox{\boldmath$k$})|^2 \rangle,  
\end{eqnarray}
where ${\hat I}_{m}(\mbox{\boldmath$k$})=\sum_{e} A_e {\exp}({-i{\mbox{\boldmath$k$}}\cdot{\mbox{\boldmath$R$}_e^{G}}}) I_m^e$ ($m$=1,2) and ${\mbox{\boldmath$R$}_e^{G}}=(1/3)\sum_{i=1,2,3}{\mbox{\boldmath$R$}_i}$ is the geometrical center of the $e$-th triangular element. 
Note that, although a window function is not used here, the resultant structure factor for $k>2\times 2\pi/L$ is hardly altered by whether implementing it or not. 
In the continuum limit, it may be reduced to ${\hat I}_{m}(\mbox{\boldmath$k$})=\int d{\mbox{\boldmath$r$}} {\exp}({-i{\mbox{\boldmath$k$}}\cdot{\mbox{\boldmath$r$}}}) w({\mbox{\boldmath$r$}}) I_m({\mbox{\boldmath$r$}})$. 
As denoted in the main text, anisotropy and localization near the walls of the particle motions are not remarkable for the present ranges of $\phi$ and $\dot\gamma$, and thus we may safely replace $S_{I_m}( { {\mbox{\boldmath$k$}}})$ by its angle average $S_{I_m}(k)$. 

Figures \ref{FigI1} and \ref{FigI2} display $S_{I_1}(k)$ and $S_{I_2}(k)$, respectively.   
As demonstrated in Fig. 5 in the main text, the regions with larger values of $I_2$ and $i_2$ almost correspond to each other, which indicates that stronger solvent dissipation is associated with larger relative shearing particle motion.  
This correspondence is further supported by Fig. \ref{FigI2}: the spatial correlation of $I_2$ and its intensity grow as $\phi$ increases and $\dot\gamma$ decreases. Such tendencies of $S_{I_2}(k)$ are qualitatively similar to those of $S(k)$, the structure factor of $i_2$, which is shown in Fig. 4 in the main text. 
Unlike $S_{I_2}(k)$, Fig. \ref{FigI1} shows that an enhancement of $S_{I_1}(k)$ at the lower $k$ is highly suppressed: as indicated in the main text, particle density increases and decreases occur alternatively in space at several particle length scales, which may suppress $I_1$ at larger scales. 
Since the rearrangement dynamics of particles are inevitably accompanied by solvent flows surrounding them, it is necessary to form flow channels by enhancing the particle gaps. However, the incompressibility of the solvent components tends to prevent particle density changes. The obtained form of $S_{I_1}(k)$ may reflect a compromise of these contradicting requirements.


\begin{thebibliography}{70}
\bibitem{comment_anisotropy}
{ When an isolated circular particle is put under simple shear flow, $i_2$ around the particle exhibits four-fold symmetry, creating square-like patterns around the particle domain. Even in semi-dilute regions, the $i_2$ field is supposed to be approximated as a superposition of contributions from each particle, resulting in square-like (dark-colored) patterns around the particle regions. This effect is emphasized by the color tones selected for Fig. 2. }
\bibitem{comment_anisotropic}
Particle overlaps (gaps) are more enhanced than in stationary states along the compression (extension) axis, whereby the particle-size reduction may occur anisotropically \cite{Furukawa23}.
\bibitem{comment_overlap}
Increasing $\dot\gamma$ enhances particle overlap mainly along the compression axis ($\hat x=-{\hat y}$). At the same time, mainly along the elongation axis (${\hat x}={\hat y}$), the gap between particles is also increased to a similar degree.  
Note, however, that our preliminary simulations show that the distortion of the particle configuration is not exactly symmetric along the compression and elongation axes and that the average interparticle distance becomes smaller with increasing $\dot\gamma$.
\end{thebibliography}
\end{document}